\documentclass[aps,prb,floatfix,reprint,twocolumn,reprint,amsmath,amssymb,superscriptaddress]{revtex4}
\usepackage{graphicx}
\usepackage{bm}
\usepackage{amsmath}
\usepackage{dcolumn}
\usepackage{dsfont}
\usepackage{amssymb}
\usepackage{tabularx}
\usepackage{array}
\usepackage{float}
\usepackage{color}
\usepackage{epstopdf}
\usepackage{mathrsfs}
\usepackage[colorlinks, linkcolor=blue,anchorcolor=blue,citecolor=blue,urlcolor=blue]{hyperref}

\begin{document}
\title{Energy-lowering symmetry breaking creates a flat-band insulator in paramagnetic Nb$_3$Cl$_8$}

\author{Jia-Xin Xiong}
\affiliation{Renewable and Sustainable Energy Institute, University of Colorado, Boulder, Colorado 80309, USA}
\author{Xiuwen Zhang}
\affiliation{Renewable and Sustainable Energy Institute, University of Colorado, Boulder, Colorado 80309, USA}
\author{Alex Zunger}
\email{alex.zunger@colorado.edu}
\affiliation{Renewable and Sustainable Energy Institute, University of Colorado, Boulder, Colorado 80309, USA}

%\date{\today}
\begin{abstract}
Ordinary band structure calculations of quantum materials often incorrectly predicted metallic, instead of insulating electronic structure, motivating Mott-Hubbard strong electron correlation as a gapping mechanism. More recently, allowing the formation of local structural symmetry breaking motifs in otherwise ordinary band theory was shown to lower the total energy while predicting insulating gaps when they are observed. An important counter example was recently pointed out whereby the flat band formed in Nb$_3$Cl$_8$ by symmetry breaking is by itself partially occupied, thus failing to account for the observed insulating state. It is shown here that a generalized symmetry breaking involving the cooperative structural and magnetic effects produces in mean-field-like density functional theory an energy lowering insulating phase.
\end{abstract}

\maketitle
%\section{Introduction}
%\emph{{\color{blue}Gap anomaly in Nb$_3$Cl$_8$}:}
\textit{Gap anomaly in Nb$_3$Cl$_8$:} A range of quantum materials manifest local microscopic degrees of freedom nested within the average crystallographic structure. Such local motifs are experimentally observable by local structural probes [1,2] and calculable quantum mechanically as energy-lowering symmetry breaking motifs [3,4]. Symmetry breaking can split off narrow flat bands [5–8] from a broad conduction band (as in LaTiO$_3$ and VO$_2$), or from the broad valence band (as in LaFeO$_3$ and SrBiO$_3$), leading in a single step to the observed insulating phases in these quantum oxides. This offers a link between energy gap formation and the local structural motifs  traditional in solid-state chemistry, such as Peierls instabilities, Jahn-Teller distortions, or bond disproportionation, both in d-electron (such as LaMnO$_3$ [9] and NbO$_2$ [10]) or in non-d-electron compounds (such as BaBiO$_3$ [11]). It was recently pointed out [12,13] that $\alpha$-phase Nb$_3$Cl$_8$ is special among other quantum materials in that symmetry breaking (here, Nb-Nb-Nb trimer formation) creates a partially occupied band in the ground state, hence the system remains metallic whereas the observed $\alpha$-phase [12,13] is insulating. This discrepancy opens the possibility that $\alpha$-phase Nb$_3$Cl$_8$ (see Supplemental Section A on the structures) is the long-sought example where purely electronic strong correlation, encoded by the Mott-Hubbard approach [14] in a fixed crystallographic structure, is not just a textbook illustration, but is required for explaining gaping unobtainable by uncorrelated physics.

\textit{The significance of this discovery:} This triumph [12] is far from being trivial. It requires avoidance of numerous alternatives explainable via conventional band structure such as density functional theory (DFT). Had the composition of the measured sample been Cl-richer (such as Nb$_3$Cl$_9$) or one where one Cl atom would be replaced by Te (such as in Nb$_3$Cl$_7$Te [15]), the correct insulating phases would be guaranteed by the trivial closed-shell configuration. Structural and magnetic symmetry breaking must be examined, although they were rejected to establish a compelling alternative scenario for insulation via strong correlation. For example, the sister d-electron compound NbO$_2$ is predicted by DFT to be a metal without any structural symmetry breaking, but once Peierls-like symmetry breaking of Nb-Nb dimerization is examined [16], it shows total energy stabilization, concomitantly with becoming an insulator, in agreement with experiments without need for strong correlation Hubbard description [Figure 1(a, b)]. Also, other perovskites like SrVO$_3$ [17] and LaNiO$_3$ [18] are predicted by DFT to be metals, and structural or magnetic symmetry breaking is too weak to convert them to insulators. However, the predicted metallic states in the former oxides are the correct experimental reality for these two compounds [19–21], unlike the case for the insulating $\alpha$-phase of Nb$_3$Cl$_8$.

\textit{The question posed:} Can symmetry breaking—either structural or magnetic, or both—lower the total energy, while also opening a band gap? Here we examine the possible role of magnetism in explaining the insulation of Nb$_3$Cl$_8$. We use the DFT exchange-correlation functional [22] meta-generalized gradient approximation (meta-GGA) to calculate the electronic structures without Coulomb repulsion $U$. The DFT method details are provided in Supplemental Section B.

\begin{figure}[!ht]
	\centering
	\includegraphics[width=1.0\linewidth]{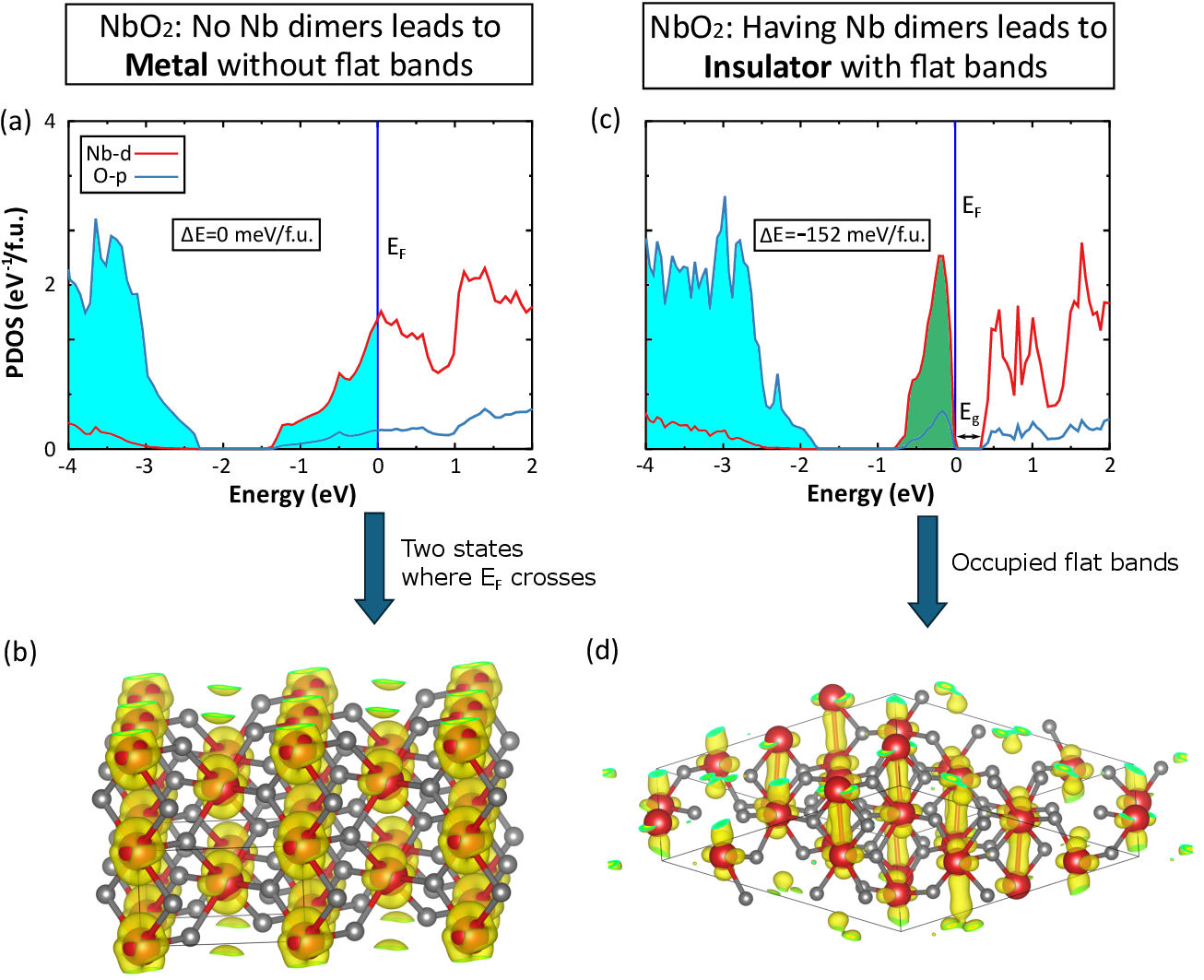}
	\caption{Two phases of nonmagnetic tetragonal NbO$_2$. (a, b) The metallic phase without Nb dimers and (c, d) the insulating phase with Nb dimers creating occupied flat bands. (a, c) The partial densities of states (PDOS) decomposed into Nb-d (red) and O-p (blue) orbitals. The filled blue and green colors in (a, c) represent the occupied states. The filled green color in (c) emphasizes the occupied split-off flat bands in the insulating phase. (b, d) The partial charge density distributions for the states (b) near the Fermi level and (d) occupied flat bands. The partial charge densities use the isosurface of 0.035 and 0.025 |e|/bohr$^3$, respectively. The dark red and grey balls represent Nb and O atoms, respectively.}
	\label{fig1}
\end{figure}

\begin{figure*}[!htb]
\centering
\includegraphics[width=1.0\linewidth]{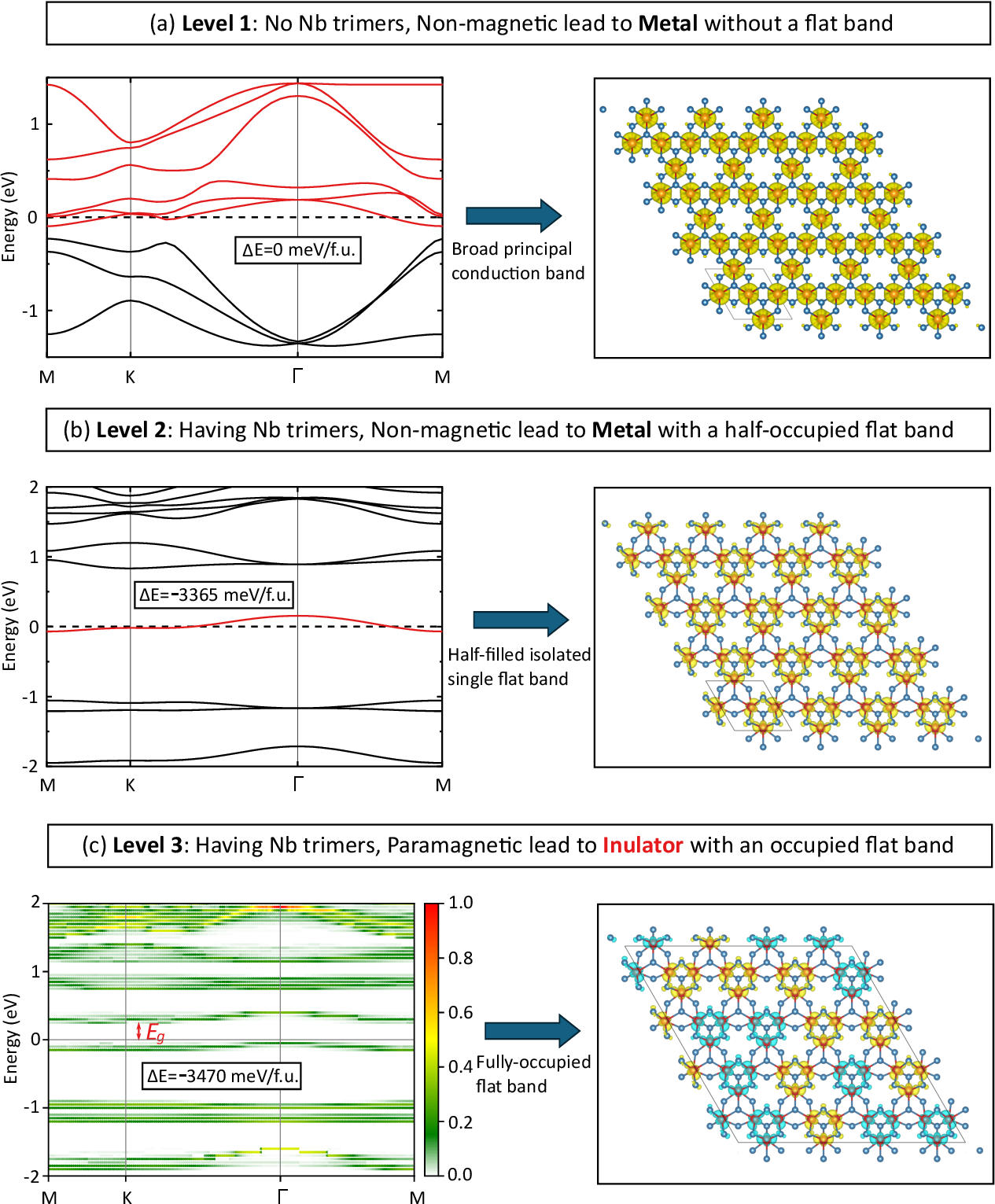}
\caption{Three levels of monolayer Nb$_3$Cl$_8$ including the band structures (left panels) and (magnetic) partial charge densities near the Fermi level (right panels). (a) Level 1: Non-magnetic phase (1 f.u. unit cell) having no Nb trimers leads to metal without a flat band. (b) Level 2: Non-magnetic phase (1 f.u. unit cell) having Nb trimers leads to metal with an isolated half-filled flat band marked in red in the band structure. (c) Level 3: Paramagnetic phase having Nb trimers leads to insulator. The unfolded band structure is obtained by projecting the states of 16 f.u. supercell onto primitive cell with the color bar representing the spectral function. The yellow and shallow blue color in magnetic charge density mean positive and negative magnetic densities, respectively. The partial charge densities in (a, b) use the isosurface of 0.02 and 0.008 $|e|$/bohr$^3$, respectively, and the magnetic density in (c) uses the isosurface of 0.004 $|e|$/bohr$^3$. The red and blue balls represent Nb and Cl atoms, respectively.}
\label{fig2}
\end{figure*}

\textit{Structural symmetry breaking in nonmagnetic NbO$_2$ suffices to correctly describe it as an insulator:} Before considering Nb$_3$Cl$_8$, we first glance again at its simpler sister compound NbO$_2$. This compound has a metal-insulator transition with two nonmagnetic phases whose crystal structures are both tetragonal but different in formation of Nb-Nb dimers. As shown in Fig. 1(a, b), the high-temperature (high-$T$) phase has the Fermi level residing in the broad principal conduction band and is metallic because it has no Nb-Nb dimers [Fig. 1(a)]. As shown in Fig. 1(c, d) the low-temperature (low-$T$) phase has Nb-Nb dimerization [Fig. 1(c)]. This dimerization constitutes structural symmetry breaking which stabilizes the total energy by 152 meV/formula unit relative to the metallic phase and splits off occupied flat bands from the principal conduction band down into the internal energy gap, thereby opening a band gap [Fig. 1(b)]. The states of the split-off flat bands in the low-$T$ insulating phase are mainly localized on the Nb-Nb dimers [Fig. 1(d)], while the states near the Fermi level in the high-$T$ metallic phase surround the Nb atoms [Fig. 1(b)].

\textit{The structural symmetry breaking in Nb$_3$Cl$_8$ retains the incorrect metallic phase:} The compound Nb$_3$Cl$_8$ is more complex than its sister compound NbO$_2$ in terms of the structural symmetry breaking, paramagnetism and thus band gap opening mechanism. Nb$_3$Cl$_8$ is a layered material where each layer has Nb trimers forming the distorted Kagome lattice with a breathing mode [12,23]. Each Nb trimer is surrounded by 13 Cl ions, forming an Nb$_3$Cl$_{13}$ cluster [Nb$_3$Cl$_8$]$^{5+}$[Cl$_5$]$^{5-}$. Here a Nb trimer has 15 electrons but 8 of them are transferred to Cl$_8$, so 7 electrons are left per cluster. They occupy the molecular orbitals configuration (1a$_1$)$^2$(1e)$^4$(2a$_1$)$^1$ so that the highest occupied molecular orbital (HOMO) is singly occupied for the nonmagnetic Nb trimer cluster. By comparing the band structures of monolayer and bilayer structures for the $\alpha$-phase Nb$_3$Cl$_8$, Supplemental Section C shows that the interlayer coupling does not influence the metallic state of the non-magnetic phase. Thus, subsequent calculations are performed by using the monolayer structure. The Nb-Nb-Nb trimerization in $\alpha$-phase Nb$_3$Cl$_8$ is observed experimentally, showing short intracluster Nb-Nb bond length $d^{short}(Nb-Nb)$ of 2.834 \AA and long intercluster Nb-Nb bond length $d^{long}(Nb-Nb)$ of 3.998 \AA [24]. The DFT structural relaxed calculations giving intracluster distance of 2.85 \AA and intercluster of 4.02 \AA are close to the experimental measurement [24]. We define the “degree of trimerization” (DOT) as:
\begin{equation}
DOT=\frac{d^{long}(Nb-Nb)-d^{short}(Nb-Nb)}{d^{long}(Nb-Nb)+d^{short}(Nb-Nb)}\times 100\%.
\end{equation}
The DOT for the DFT relaxed structure and for the  experimental structure [24] are both estimated to be 17.0\%. The calculated total energy gives an enormous stabilization of 3365 meV/f.u. but the system is a band structure metal.

\textit{Paramagnetism in $\alpha$-phase Nb$_3$Cl$_8$:} The paramagnetic (PM) $\alpha$-phase of Nb$_3$Cl$_8$ was observed experimentally by measuring the magnetic susceptibility as a function of temperature, which was well fitted to the Curie-Weiss function giving rise to an effective PM Bohr magneton of 1.65 $\mu_B$ [12]. This paramagnetism is not considered relevant to gapping in the strongly correlated approach. In such Mott model, magnetic exchange J deciding magnetism is modeled perturbatively in terms of Mott-ness strong Coulomb repulsion $J\verb|~|$1/$U$, leading to the expected weak or no magnetism when a large Coulomb $U$ is selected. Here, we do not exclude the PM configuration.

\textit{Description of paramagnetic symmetry breaking:} We compute PM non-perturbatively using magnetic exchange interactions $J$ obtained from spin polarized DFT. The PM configuration is not replaced a priori by the nonmagnetic structure where each site has zero moment. The latter view assumes that the local symmetry equals the average crystal structure of global symmetry. Instead, a PM configuration can be described more generally as a distribution of non-zero local magnetic moment. We use for this purpose a “specially constructed” [25] supercell consisting of $n$ Nb$_3$Cl$_8$ formula units (where $n$ is a size convergence parameter and we use $n$=16 here) with zero global magnetization. Much like “special $k$-points for Brillouin zone integrating” [26,27], the spin-special quasirandom structure (spin-SQS) method is not a single snapshot but represents a configurational average. Indeed, the spin-SQS searches for a magnetic configuration that best matches the target multisite spin-spin correlation function with the given short-range order. The local magnetic moments are decided by DFT energy minimization and charge self-consistency starting from the optimal spin-SQS magnetic structure. Since the band structure of PM phase in a small Brillouin zone (corresponding to a real space supercell) is inconveniently dense to observe important information such as band dispersion, we unfold the band structure into the original Brillouin zone thereby obtaining effective spectral functions [28] that have, unlike conventional band structure, natural energy-dependent intensities and “fuzziness”.

\textit{The important observation regarding magnetic symmetry breaking:} Once structural symmetry breaking (Nb trimerization) is accounted for, we find that magnetic symmetry breaking opens a gap in the PM phase, explaining the insulating phase without recourse to strong correlation or to dynamic effects. This is described next by breaking the process into steps that clarify the individual contributions.

\textit{Hierachy of symmetry breaking:} Figure 2 demonstrates the three successive levels of theoretical description of monolayer Nb$_3$Cl$_8$ based on the existence or absence of two symmetry breaking modes—Nb trimerization and magnetic symmetry breaking.

\begin{figure*}[!htb]
\centering
\includegraphics[width=1.0\linewidth]{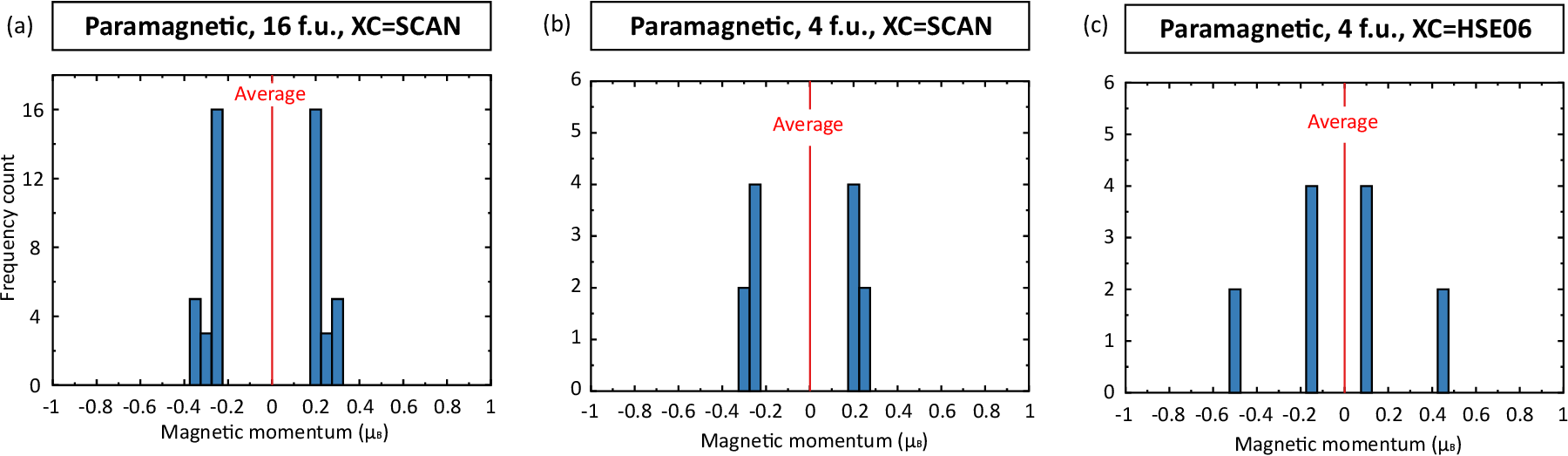}
\caption{Distribution of local magnetic moments in paramagnetic monolayer Nb$_3$Cl$_8$ simulated by different sizes of supercells using different exchange correlation (XC) functionals: (a) 16 formula-unit supercell by SCAN functional, (b) 4 formula-unit supercell by SCAN functional, and (c) 4 formula-unit supercell by HSE06 functional. The red solid lines at zero magnetic moment represent the average magnetic moment to be zero. The bin width of magnetic momentums for plot is 0.05 $\mu_B$.}
\label{fig3}
\end{figure*}

\textit{Level 1} (symmetry unbroken) is for the hypothetical structure without Nb trimerization and magnetism. As shown in Fig. 2(a), the Fermi level lies within a broad principal conduction band with its charge density distributed equally on the Nb atoms. This configuration is taken as the total energy reference $\Delta E$=0.

In \textit{Level 2} (structural symmetry breaking) with the Nb trimerization, we find that the Nb trimerization stabilizes the total energy by 3365 meV/f.u. relative to Level 1 and induces a single half-occupied isolated flat band at the Fermi level. This isolated flat band with band width of $\verb|~|$0.3 eV emerges from the broad principal conduction band of Level 1 and its charge density is dominantly located around the Nb trimers, as shown in Fig. 2(b). For the bilayer structure of nonmagnetic $\alpha$-phase Nb$_3$Cl$_8$, there are two split-off flat bands crossing with the Fermi level that are degenerate at H point (See Supplemental Section B). Evidently, in both monolayer and bilayer structures of $\alpha$-phase Nb$_3$Cl$_8$, the Nb trimerization with the formation of partially occupied split-off flat bands cannot open a band gap.

In \textit{Level 3} (structural and magnetic symmetry breaking) that considers both (i) Nb trimers due to structural symmetry breaking and (ii) magnetic symmetry breaking. The latter is depicted as PM spin-SQS with self-consistently determined distribution of local moments, as shown in Fig. 2(c). This shows that the half-filled isolated flat band of Level 2 split forming two isolated flat bands and creating a band gap of 0.3 eV. It increases to 1.2 eV when using the HSE06 exchange correlation functional instead of SCAN. This process from Level 2 to Level 3 reduces the total energy by 105 meV/f.u. The band gap is opened by interaction of (i) structural symmetry breaking (forming trimers) and (ii) magnetic symmetry breaking. We find that the band gap cannot be opened if there is only one of the two symmetry breaking modes. This shows that one can predict the insulating phase of such quantum materials without Hubbard $U$.

\textit{Distribution of local motifs:} To further indicate the difference between PM and nonmagnetic phases, Figure 3 shows the distribution of local magnetic moments in PM Nb$_3$Cl$_8$ (Level 3). It is  observed that: (i) All of these results present a distribution of non-zero local magnetic moments and an average zero net magnetic moment, which is independent of the exchange correlation functionals and supercell sizes; (ii) The distribution of magnetic moments for SCAN and HSE06 is very different partially due to the strong exchange interaction in HSE06, but the choice of different exchange correlation functionals do not qualitatively change our main point that PM is essential to open the band gap.

\textit{Discussion:} The PM phase lacks the long-range order of the AFM phase but has short-range order that can still be essential to open the band gap. Indeed long-range antiferromagnetic order is well known [29] to enable insulating gaps. However, the absence of long-range magnetic order as the case in PM phase can, in the presence of short-range order lead either (i) to metallic phase if symmetry breaking is weak, as in SrVO$_3$ [3,17] and LaNiO$_3$ [30,31], or (ii) to an insulating PM phase, as in MnO, FeO, CoO, and NiO. Furthermore, symmetry breaking can induce other effects such as effective mass enhancement [17] without strong correlation.

The present results indicate that approximating the PM as nonmagnetic configuration and using primitive unit cell as input to do the calculations is not appropriate, missing energy lowering and gapping. Compared to nonmagnetic NbO$_2$ that only needs structural symmetry breaking (Nb-Nb dimerization) to open a band gap (Fig. 1), $\alpha$-phase Nb$_3$Cl$_8$ is more complicated because it needs both structural (Nb-Nb-Nb trimerization) and magnetic (PM) symmetry breaking modes (Fig. 2).

\textit{The insulating bilayer nonmagnetic $\beta$-phase Nb$_3$Cl$_8$:} Besides the high-$T$ PM $\alpha$-phase, Nb$_3$Cl$_8$ has a low-$T$ nonmagnetic $\beta$-phase, which is measured also as an insulator [12]. Because of the proximity of the layers in the $\beta$-phase, we allow in this case interlayer interaction via adding the Van der Waals (VdW) dispersive forces to the regular DFT. These dispersive forces have no influence on the $\alpha$-phase but shorten the interlayer distance and increase the coupling within the bilayer $\beta$-phase. We find that in nonmagnetic $\beta$-phase Nb$_3$Cl$_8$ with two chemical formulae in the primitive cell, there are two split-off flat bands crossing with the Fermi level that are not degenerate at any $k$-points (see Supplemental Section C). This energy splitting is induced by the interlayer interaction. Therefore, in the $\beta$-phase the Nb trimerization and coupling of Nb trimers from adjacent layers have the potential to open a band gap, even without magnetism or strong correlation. Supplemental Section C provides a few tests of the insulating gap under different crystal structural parameters. A detailed study of the insulating gap is $\beta$-phase Nb$_3$Cl$_8$ is however beyond our scope as the details of the interlayer geometry are unknown.

Many-body physics is certainly to be preferred over single particle theory for understanding multiplet effects in electronically isolated ions in the gas phase [32,33], or for $d$-electron ions in wide-gap solids [34,35], etc. Many-body effects certainly exist as spectroscopic multiplet effects [36,37] and in multi excitons. Still, multi-electron effects that are often apparent in isolated ions such as carbon [38] often “melt away” when such atoms bond in condensed phases—as is the case when diamond or graphite are formed from carbon atoms. So, the question is: when is a many-electron treatment unescapable? But we see that the outlier phenomenology of gapping, the unusual case of Nb$_3$Cl$_8$ is describable by conventional DFT while reproducing also observable structural (Nb-Nb-Nb trimerization) and magnetic (the stabilization of a PM) symmetry breaking. This leads to an interesting duality: How will the band gap change if both the observable symmetry breaking (discussed in the present work) and the strong correlation Mott-Hubbard treatment used successfully in Ref. [12,23] to obtain gapping (without magnetic symmetry breaking) have important influences in producing an insulating phase? This remained question will inspire future research.

\begin{acknowledgments}
    We thank Dr. Shunye Gao for discussions of his paper denoted here as Ref. [12]. This work was supported by the U.S. Department of Energy, Office of Science, Basic Energy Sciences, Materials Sciences and Engineering Division under Grant No. DE-SC0010467. This work used resources from the National Energy Research Scientific Computing Center (NERSC), which was supported by the Office of Science of the U.S. Department of Energy.
\end{acknowledgments}

%\bibliography{ref}
\section*{Reference}
\noindent[1]	A. N. Beecher, O. E. Semonin, J. M. Skelton, J. M. Frost, M. W. Terban, H. Zhai, A. Alatas, J. S. Owen, A. Walsh, and S. J. L. Billinge, Direct observation of dynamic symmetry breaking above room temperature in methylammonium lead iodide perovskite, ACS Energy Lett. 1, 880 (2016).

\noindent[2]	E. S. Božin, C. D. Malliakas, P. Souvatzis, T. Proffen, N. A. Spaldin, M. G. Kanatzidis, and S. J. L. Billinge, Entropically stabilized local dipole formation in lead chalcogenides, Science 330, 1660 (2010).

\noindent[3]	O. I. Malyi and A. Zunger, False metals, real insulators, and degenerate gapped metals, Appl. Phys. Rev. 7, 041310 (2020).

\noindent[4]	X.-G. Zhao, Z. Wang, O. I. Malyi, and A. Zunger, Effect of static local distortions vs. dynamic motions on the stability and band gaps of cubic oxide and halide perovskites, Materials Today 49, 107 (2021).

\noindent[5]	Z. Zhu, H. Peelaers, and C. G. Van De Walle, Electronic and protonic conduction in LaFeO3, J. Mater. Chem. A 5, 15367 (2017).

\noindent[6]	M. Eaton, A. Catellani, and A. Calzolari, VO2 as a natural optical metamaterial, Opt. Express 26, 5342 (2018).

\noindent[7]	F. El-Mellouhi, E. N. Brothers, M. J. Lucero, I. W. Bulik, and G. E. Scuseria, Structural phase transitions of the metal oxide perovskites SrTiO3, LaAlO3, and LaTiO3 studied with a screened hybrid functional, Phys. Rev. B 87, 035107 (2013).

\noindent[8]	J. Varignon, J. Santamaria, and M. Bibes, Electrically switchable and tunable Rashba-type spin splitting in covalent perovskite oxides, Phys. Rev. Lett. 122, 116401 (2019).

\noindent[9]	H. Sawada, Y. Morikawa, K. Terakura, and N. Hamada, Jahn-Teller distortion and magnetic structures in LaMnO3, Phys. Rev. B 56, 12154 (1997).

\noindent[10]	A. O’Hara and A. A. Demkov, Nature of the metal-insulator transition in NbO2, Phys. Rev. B 91, 094305 (2015).

\noindent[11]	S. Sarkar, R. Raghunathan, S. Chowdhury, R. J. Choudhary, and D. M. Phase, The mystery behind dynamic charge disproportionation in BaBiO3, Nano Lett. 21, 8433 (2021).

\noindent[12]	S. Gao, S. Zhang, C. Wang, S. Yan, X. Han, X. Ji, W. Tao, J. Liu, T. Wang, S. Yuan, G. Qu, Z. Chen, Y. Zhang, J. Huang, M. Pan, S. Peng, Y. Hu, H. Li, Y. Huang, H. Zhou, S. Meng, L. Yang, Z. Wang, Y. Yao, Z. Chen, M. Shi, H. Ding, H. Yang, K. Jiang, Y. Li, H. Lei, Y. Shi, H. Weng, and T. Qian, Discovery of a single-band Mott insulator in a van der Waals flat-band compound, Phys. Rev. X 13, 041049 (2023).

\noindent[13]	Y. Haraguchi, C. Michioka, M. Ishikawa, Y. Nakano, H. Yamochi, H. Ueda, and K. Yoshimura, Magnetic–nonmagnetic phase transition with interlayer charge disproportionation of Nb3 trimers in the cluster compound Nb3Cl8, Inorg. Chem. 56, 3483 (2017).

\noindent[14]	J. Hubbard and B. H. Flowers, Electron correlations in narrow energy bands, Proc. R. Soc. A 276, 238 (1963).

\noindent[15]	H. Zhang, Z. Shi, Z. Jiang, M. Yang, J. Zhang, Z. Meng, T. Hu, F. Liu, L. Cheng, Y. Xie, J. Zhuang, H. Feng, W. Hao, D. Shen, and Y. Du, Topological flat bands in 2D breathing-Kagome lattice Nb3TeCl7, Adv. Mater. 35, 2301790 (2023).

\noindent[16]	M. J. Wahila, G. Paez, C. N. Singh, A. Regoutz, S. Sallis, M. J. Zuba, J. Rana, M. B. Tellekamp, J. E. Boschker, T. Markurt, J. E. N. Swallow, L. A. H. Jones, T. D. Veal, W. Yang, T.-L. Lee, F. Rodolakis, J. T. Sadowski, D. Prendergast, W.-C. Lee, W. A. Doolittle, and L. F. J. Piper, Evidence of a second-order Peierls-driven metal-insulator transition in crystalline NbO2, Phys. Rev. Mater. 3, 074602 (2019).

\noindent[17]	Z. Wang, O. I. Malyi, X. Zhao, and A. Zunger, Mass enhancement in 3d and s-p perovskites from symmetry breaking, Phys. Rev. B 103, 165110 (2021).

\noindent[18]	G. Gou, I. Grinberg, A. M. Rappe, and J. M. Rondinelli, Lattice normal modes and electronic properties of the correlated metal LaNiO3, Phys. Rev. B 84, 144101 (2011).

\noindent[19]	L. Zhang, Y. Zhou, L. Guo, W. Zhao, A. Barnes, H.-T. Zhang, C. Eaton, Y. Zheng, M. Brahlek, H. F. Haneef, N. J. Podraza, M. H. W. Chan, V. Gopalan, K. M. Rabe, and R. E.-Herbert, Correlated metals as transparent conductors, Nat. Mater. 15, 2 (2016).

\noindent[20]	Y. Tomioka, T. Ito, E. Maruyama, S. Kimura, and I. Shindo, Magnetic and electronic properties of single crystals of perovskite nickelate oxide LaNiO3 prepared by the laser diode floating zone method, J. Phys. Soc. Jpn. 90, 034704 (2021).

\noindent[21]	B. Li, D. Louca, S. Yano, L. G. Marshall, J. Zhou, and J. B. Goodenough, Insulating pockets in metallic LaNiO3, Adv. Electron. Mater. 2, 1500261 (2016).

\noindent[22]	J. Sun, A. Ruzsinszky, and J. P. Perdew, Strongly constrained and appropriately normed semilocal density functional, Phys. Rev. Lett. 115, 036402 (2015).

\noindent[23]	S. Grytsiuk, M. I. Katsnelson, E. G. C. P. van Loon, and M. Rösner, Nb3Cl8: A prototypical layered Mott-Hubbard insulator, npj Quantum Mater. 9, 1 (2024).

\noindent[24]	M. Ströbele, J. Glaser, A. Lachgar, and H.-J. Meyer, Structure and electrochemical study of Nb3Cl8, Z. Anorg. Allg. Chem. 627, 2002 (2001).

\noindent[25]	A. Zunger, S.-H. Wei, L. G. Ferreira, and J. E. Bernard, Special quasirandom structures, Phys. Rev. Lett. 65, 353 (1990).

\noindent[26]	A. Baldereschi, Mean-value point in the Brillouin zone, Phys. Rev. B 7, 5212 (1973).

\noindent[27]	D. J. Chadi and M. L. Cohen, Special points in the Brillouin zone, Phys. Rev. B 8, 5747 (1973).

\noindent[28]	V. Popescu and A. Zunger, Extracting E versus k effective band structure from supercell calculations on alloys and impurities, Phys. Rev. B 85, 085201 (2012).

\noindent[29]	J. C. Slater, Magnetic effects and the Hartree-Fock equation, Phys. Rev. 82, 538 (1951).

\noindent[30]	G. Catalan, Progress in perovskite nickelate research, Phase Transitions 81, 729 (2008).

\noindent[31]	J. Varignon, M. N. Grisolia, J. Íñiguez, A. Barthélémy, and M. Bibes, Complete phase diagram of rare-earth nickelates from first-principles, npj Quantum Mater. 2, 1 (2017).

\noindent[32]	C. E. Moore, Atomic Energy Levels as Derived from the Analyses of Optical Spectra: The Spectra of Hydrogen, Deuterium, Tritium, Helium, Lithium, Beryllium, Boron, Carbon, Nitrogen, Oxygen, Fluorine, Neon, Sodium, Magnesium, Aluminum, Silicon, Phosphorus, Sulfur, Chlorine, Argon, Potassium, Calcium, Scandium, Titanium, and Vanadium (U.S. Department of Commerce, National Bureau of Standards, 1949).

\noindent[33]	C. E. Moore, A Multiplet Table of Astrophysical Interest (U.S. Department of Commerce, Office of Technical Services, 1959).

\noindent[34]	M. Imada, A. Fujimori, and Y. Tokura, Metal-insulator transitions, Rev. Mod. Phys. 70, 1039 (1998).

\noindent[35]	A. Fazzio and A. Zunger, Many-electron multiplet effects in the optical spectra of NiO, CoO and MnO, Solid State Commun. 52, 265 (1984).

\noindent[36]	A. Zunger, Electronic structure of 3d Transition-Atom Impurities in Semiconductors, in Solid State Phys., edit. F. Seitz, D. Tumbull, and H. Ehrenreich (Acad Press NY) 39, 275-464 (1986).

\noindent[37]	J. S. Griffith, The Theory of Transition-Metal Ions (Cambridge University Press, 1971).

\noindent[38]	G. M. Lawrence and B. D. Savage, Radiative lifetimes of UV multiplets in boron, carbon, and nitrogen, Phys. Rev. 141, 67 (1966).

\end{document}

\title{Supplemental Material to "Energy-lowering symmetry breaking creates a flat-band insulator in paramagnetic Nb$_3$Cl$_8$"}

\author{Jia-Xin Xiong}
\affiliation{Renewable and Sustainable Energy Institute, University of Colorado, Boulder, Colorado 80309, USA}
\author{Xiuwen Zhang}
\affiliation{Renewable and Sustainable Energy Institute, University of Colorado, Boulder, Colorado 80309, USA}
\author{Alex Zunger}
\email{alex.zunger@colorado.edu}
\affiliation{Renewable and Sustainable Energy Institute, University of Colorado, Boulder, Colorado 80309, USA}

%\date{\today}

\maketitle
\section*{Supplemental Section A: Crystal and magnetic structures of $\alpha$-phase and $\beta$-phase Nb$_3$Cl$_8$}
%\emph{{\color{blue}Gap anomaly in Nb$_3$Cl$_8$}:}
In this Supplemental Section, we show the crystal structures of high-temperature (high-$T$) paramagnetic (PM) $\alpha$-phase and low-temperature (low-$T$) non-magnetic $\beta$-phase Nb$_3$Cl$_8$.

\begin{figure}[!htb]
\centering
\includegraphics[width=1.0\linewidth]{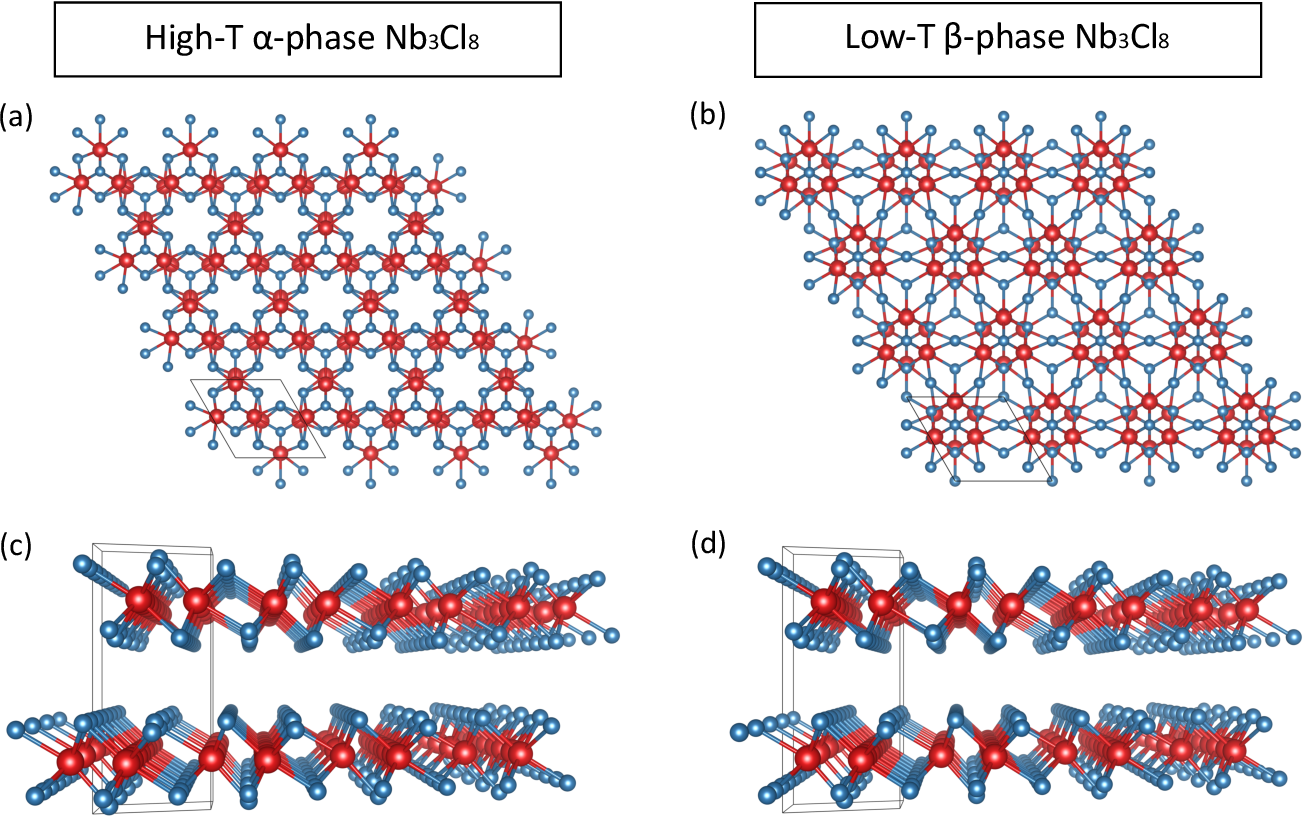}
\caption{Crystal structures of (a, c) high-temperature $\alpha$-phase and (b, d) low-temperature $\beta$-phase Nb$_3$Cl$_8$. (a, b) are the top views and (c, d) are the side views. Red and blue balls represent Nb and Cl atoms, respectively.}
\label{figS1}
\end{figure}

\section*{Supplemental Section B: DFT calculation method of electronic structures}

\begin{figure}[!htb]
\centering
\includegraphics[width=1.0\linewidth]{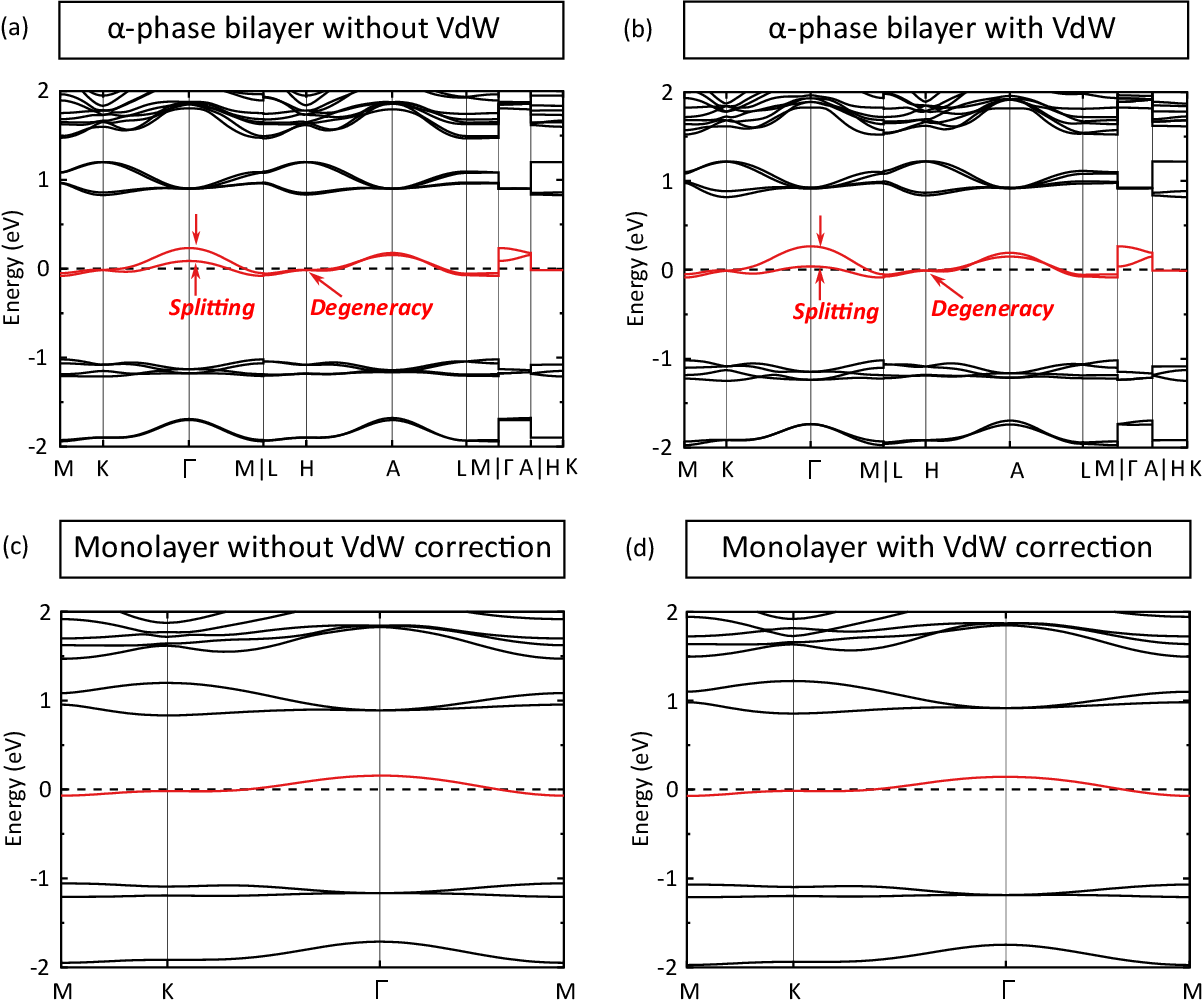}
\caption{Band structures of non-magnetic Nb$_3$Cl$_8$ of (a, b) bilayer $\alpha$-phase and (c, d) monolayer. The Van der Waals (VdW) correction is not considered in (a, c) but considered in (b, d). All the calculations use PBE+$U$ in the relaxation of crystal structures and SCAN in the static calculations. The $U$ value in PBE+$U$ is set as 2 eV. The partially occupied split-off flat bands are marked in red.}
\label{figS2}
\end{figure}

We use a plane-wave pseudopotential density functional method as implemented by the Vienna Ab initio Simulation Package (VASP) [1,2] to perform first-principle calculations. For calculation of structural relaxation, we use the projector augmented-wave (PAW) method [3,4] with the Perdew-Burke-Ernzerhof (PBE) exchange correlation functional within the generalized gradient approximation (GGA) [5] supplemented by the Coulomb $U$ (being 2 eV taken from Ref. [6]). For the electronic structure calculations, we use the Strongly Constrained and Approximately Normed (SCAN) [7] meta-generalized gradient approximation (meta-GGA) functional and Heyd-Scuseria-Ernzerhof (HSE06) hybrid functional [8,9] in the main text. For HSE06 calculations, we use the mixed factor of 0.25 for the Hartree-Fock exchange with the screening parameter of 0.2 \AA$^{-1}$ to separate the short-range and long-range interactions. The PAW-PBE pseudopotentials of VASP are utilized to represent the ion cores, effectively reducing the number of explicit electrons. The valence electrons of elements in NbO$_2$ and Nb$_3$Cl$_8$ include 11 electrons in Nb (4$p^6$5$s^1$4$d^4$), 6 electrons in O (2$s^2$2$p^4$), and 7 electrons in Cl (2$s^2$2$p^5$).

Detailed parameters in the DFT calculations are: (i) The cut-off energy for the plane-wave basis is set to be 350 eV; (ii) An uniformed tolerance for total energy convergence is set to be 10-6 eV/cell; (iii) The convergence for relaxations including atomic displacements and magnetic momentum changes are set to be 10$^{-2}$ eV/\AA; (iv) The $\Gamma$-centered $k$-point samplings are large enough with the characteristic length more than 25 to guarantee the convergence of charge density. SOC is neglected in the calculations.

Supplemental Section A gives the bilayer structures of both $\alpha$- and $\beta$-phase of Nb$_3$Cl$_8$. We consider both bilayer and monolayer structures of the $\alpha$-phase Nb$_3$Cl$_8$ including nonmagnetic and paramagnetic (PM) configurations. For the monolayer structure, we construct a slab with 15 \AA thick vacuum to prevent possible interlayer interactions. In the relaxation of this slab structure, we fix the cell volume and allow the lattice and atomic positions to change.

For nonmagnetic calculations, we use the unit cell (1 formula unit) as input for full relaxation (allowing both lattice vectors and atomic positions to change) and static calculations.

For PM calculations, we enlarge the unit cell of nonmagnetic phase making a large supercell with 16 formula units and use the spin-special quasirandom structure (spin-SQS) method [10,11] to obtain the disordered spin configuration as a representative snapshot by minimizing the correlation functions of the supercell system. Then we use this supercell as input to do constrained relaxation (fixing the lattice vectors and allowing the atoms to move) and further static calculations.

\section*{Supplemental Section C: Influence of exchange correlation functionals and Van der Waals correction on electronic structures of $\alpha$-phase and $\beta$-phase Nb$_3$Cl$_8$}

\begin{figure}[!htb]
\centering
\includegraphics[width=1.0\linewidth]{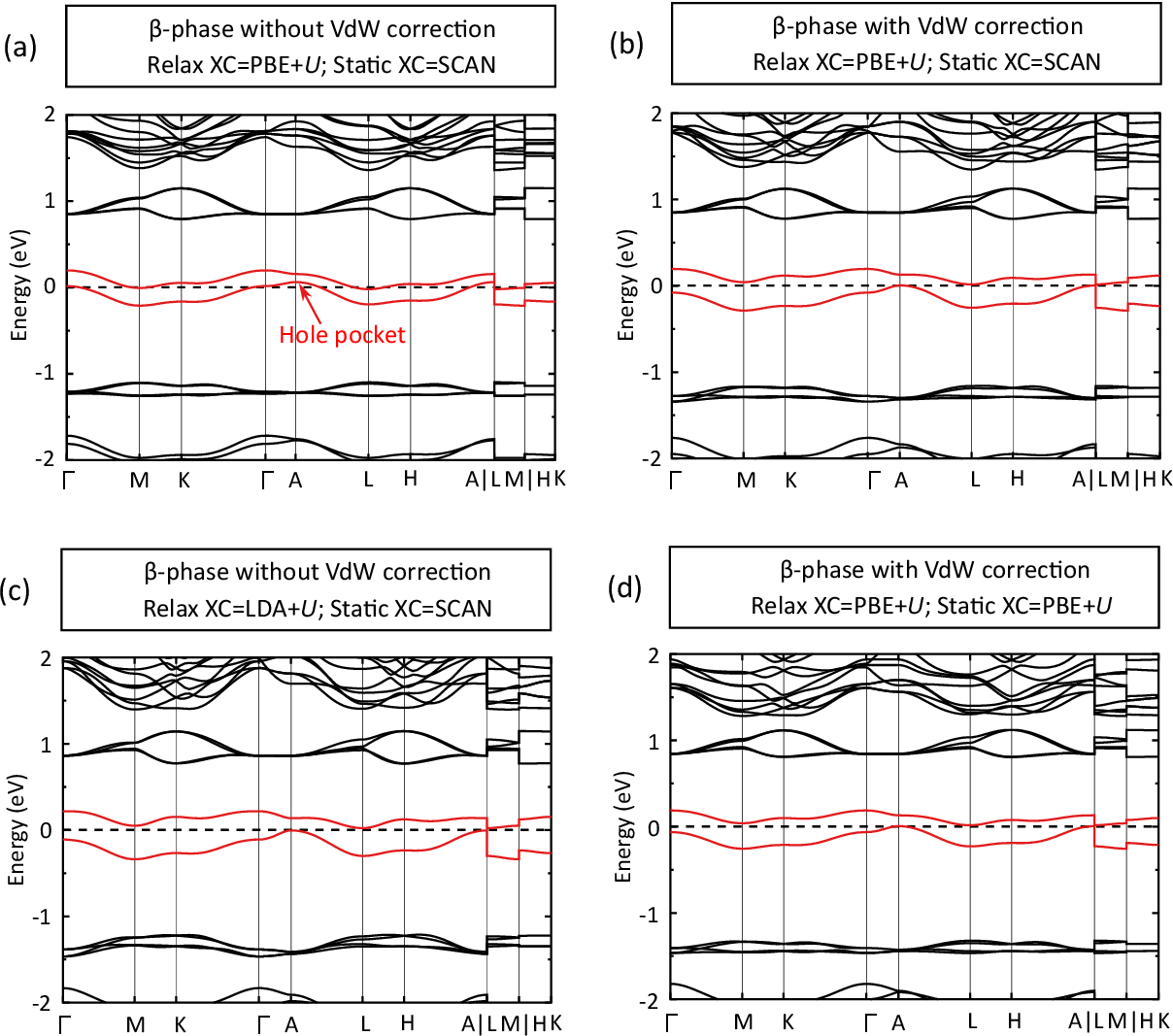}
\caption{Band structures of non-magnetic $\beta$-phase Nb$_3$Cl$_8$ calculated by different methods: (a) Without Van der Waals (VdW) correction, relaxation using PBE+$U$ and static calculation using SCAN functional. (b) With VdW correction, relaxation using PBE+$U$ and static calculation using SCAN functional. (c) Without VdW correction, relaxation using LDA+$U$ and static calculation using SCAN functional. (d) With VdW correction, relaxation and static calculation both using PBE+$U$ functional. The $U$ value in PBE+$U$ and LDA+$U$ is set as 2 eV. The two partially occupied split-off flat bands are marked in red.}
\label{figS3}
\end{figure}

In this Supplemental Section, we show the electronic structures of $\alpha$-phase (monolayer and bilayer) and $\beta$-phase Nb$_3$Cl$_8$. The differences between these two phases are (i) magnetism and (ii) different stacking mode in the perpendicular direction indicating different Nb-Nb interlayer coupling strengths. In the main text, we have investigated (i) for $\alpha$-phase. Here we ignore (i) but focus on (ii), which may be sensitive to the choice of exchange correlation functionals and allowance of Van der Waals correction. We examine the influence in the electronic structures from these two factors for both $\alpha$-phase and $\beta$-phase.

We first examine the band structure of bilayer and monolayer of $\alpha$-phase Nb$_3$Cl$_8$ [corresponding to Level 2 as shown in Figure 2(b) in the main text] as shown in Figure S2. Compared to the band structure of monolayer Nb$_3$Cl$_8$ [Fig. S2(c, d) and also Figure 2(b) in the main text], we can see that the band structure of bilayer $\alpha$-phase Nb$_3$Cl$_8$ has a twice number of bands at the same corresponding energy scales [Fig. S2(a, b)]. However, the bands are not simply folded in bilayer structure. This can be obviously observed in the two split-off flat bands with the Fermi level crossing [marked red in Fig. S2(a, b)] at the $\Gamma$ point that produce an energy splitting of 146 meV [without Van der Waals (VdW) correction as shown in Fig. S2(a)] or 226 meV [with VdW correction as shown in Fig. S2(b)]. This energy splitting reflects the small interlayer coupling in the $\alpha$-phase Nb$_3$Cl$_8$. Moreover, the states at the $H$ point are degenerated and partially occupied. This degeneracy could be protected by the $D_3$ little group symmetry at the H point, whereas the partially occupation indicates again that Nb trimerization alone cannot open the band gap.

Fig. S2 also examines the influence of VdW correction in the bilayer and monolayer of $\alpha$-phase Nb$_3$Cl$_8$. We use here the DFT-D3 method of Grimme et al [12] for the VdW correction, indicating that the VdW correction does not influence the band gap of both monolayer and bilayer of $\alpha$-phase Nb$_3$Cl$_8$ significantly except for the splitting of bands near the Fermi level, which reflects the interlayer coupling strength. This is because for $\alpha$-phase Nb$_3$Cl$_8$, both (i) Nb trimerization and (ii) paramagnetism are needed to open the band gap while the interlayer coupling is not needed. This can be clearly observed in Table S1 that DOT remains around 17.0\% for both monolayer and bilayer of $\alpha$-phase Nb$_3$Cl$_8$, though (A) the Nb-Nb interlayer distance is shortened 9.8\% by VdW correction in bilayer of $\alpha$-phase and (B) the in-plane intracluster and intercluster Nb-Nb bond lengths $d^{short}(Nb-Nb)$ and $d^{long}(Nb-Nb)$ also slightly change in both monolayer and bilayer of $\alpha$-phase.

\begin{table}[h]
    \centering
    \small
    \resizebox{\textwidth}{!}{%
    \begin{tabular}{|c | c | c | c | c |}
        \hline
         & Nb-Nb Interlayer distance(\AA) & Intracluster $d^{short}(Nb-Nb) (\AA)$ & Intercluster $d^{long}(Nb-Nb) (\AA)$ & Degree of trimerization \\ \hline
        Bilayer $\beta$-phase &  &  &  & \\ 
        No VdW; Relaxed by PBE+$U$ & 6.78 & 2.84 & 4.01 & 17.2\% \\ 
        VdW; Relaxed by PBE+$U$ & 6.24 & 2.83 & 3.97 & 16.8\% \\ 
        No VdW; Relaxed by LDA+$U$ & 6.07 & 2.80 & 3.86 & 15.9\% \\ \hline
        Bilayer $\alpha$-phase &  &  &  & \\ 
        No VdW; Relaxed by PBE+$U$ & 6.98 & 2.85 & 4.02 & 17.0\% \\ 
        VdW; Relaxed by PBE+$U$ & 6.29 & 2.83 & 3.98 & 16.9\% \\ \hline
        Monolayer $\alpha$-phase &  &  &  & \\ 
        No VdW; Relaxed by PBE+$U$ & -- & 2.85 & 4.02 & 17.0\% \\ 
        VdW; Relaxed by PBE+$U$ & -- & 2.83 & 3.99 & 17.0\% \\ \hline
    \end{tabular}%
    }
    \caption{Structure parameters relaxed by different methods for Nb$_3$Cl$_8$ of bilayer $\beta$-phase, bilayer $\alpha$-phase and monolayer $\alpha$-phase.}
    \label{tab:my_table}
\end{table}

We then examine the band structure of $\beta$-phase of Nb$_3$Cl$_8$ as shown in Figure S3. We also focus on the two split-off flat bands with the Fermi level crossing (marked in red). Compared to the nonmagnetic $\alpha$-phase Nb$_3$Cl$_8$ in Fig. S2, the states at all $k$-points are split in energy, reflecting a larger interlayer coupling strength in the $\beta$-phase. The lowest red band in Fig. S3(a) is mostly below the Fermi level with a few small hole pockets above the Fermi level around the $A$ point. This suggests a semi-metallic phase and the hole pocket around the $A$ point can be affected by the interlayer couplings and possibly by the modulated trimerization affected by interlayer couplings. There are no direct experimental measurements for the band gap of $\beta$-phase of Nb$_3$Cl$_8$, but a large band gap is almost believed and accepted by the community.

We next examine the influence of different exchange correlation functionals and VdW correction. We find that in Fig. S3(b, c) by either using a more delocalized exchange correlation functional such as local-density approximation (LDA) [13] (the degree of delocalization for different functionals: LDA$>$PBE$>$SCAN$>$HSE) or allowing the VdW correction, the hole pockets near the Fermi level around $A$ point disappears and the two flat bands tend to stay away from each other, making the system a “bad metal”. The VdW correction does not obviously influence the calculation of electronic structures based on the same crystal structure [Fig. S3(c, d)]. The change of electronic structures in Fig. S3(b, c, d) relative to that in Fig. S3(a) is because these two factors can highly determine both (i) Nb trimerization and (ii) Nb-Nb interlayer coupling which enable the band gap opening in bilayer $\beta$-phase. As shown in Table S1, (A) the Nb-Nb interlayer distance is significantly reduced by LDA functional or VdW correction, but (B) the DOT is also reduced. The competition between (A) and (B) makes the band gap still tiny in Fig. S3(b, c, d).

%\bibliography{ref}
\section*{Supplemental Reference}

\noindent[1]	G. Kresse and J. Furthmüller, Efficiency of ab-Initio total energy calculations for metals and semiconductors using a plane-wave basis set, Comp. Mater. Sci. 6, 15 (1996).

\noindent[2]	G. Kresse and J. Furthmuller, Efficient iterative schemes for ab Initio total-energy calculations using a plane-wave basis set, Phys. Rev. B 54, 11169 (1996).

\noindent[3]	P. E. Blochl, Projector augmented-wave method, Phys. Rev. B 50, 17953 (1994).

\noindent[4]	G. Kresse and D. Joubert, From ultrasoft pseudopotentials to the projector augmented-wave method, Phys. Rev. B 59, 1758 (1999).

\noindent[5]	J. P. Perdew, K. Burke, and M. Ernzerhof, Generalized gradient approximation made simple, Phys. Rev. Lett. 77, 3865 (1996).

\noindent[6]	J. Jiang, Q. Liang, R. Meng, Q. Yang, C. Tan, X. Sun, and X. Chen, Exploration of new ferromagnetic, semiconducting and biocompatible Nb3X8 (X = Cl, Br or I) monolayers with considerable visible and infrared light absorption, Nanoscale 9, 2992 (2017).
[7]	J. Sun, A. Ruzsinszky, and J. P. Perdew, Strongly constrained and appropriately normed semilocal density functional, Phys. Rev. Lett. 115, 036402 (2015).
[8]	J. Heyd and G. E. Scuseria, Hybrid functionals based on a screened Coulomb potential, J. Chem. Phys. 118, 8207 (2003).

\noindent[9]	A. V. Krukau, O. A. Vydrov, A. F. Izmaylov, and G. E. Scuseria, Influence of the exchange screening parameter on the performance of screened hybrid functionals, J. Chem. Phys. 125, 224106 (2006).

\noindent[10]	S.-H. Wei, L. G. Ferreira, J. E. Bernard, and A. Zunger, Electronic properties of random alloys: Special quasirandom structures, Phys. Rev. B 42, 9622 (1990).

\noindent[11]	A. Zunger, S.-H. Wei, L. G. Ferreira, and J. E. Bernard, Special quasirandom structures, Phys. Rev. Lett. 65, 353 (1990).

\noindent[12]	S. Grimme, J. Antony, S. Ehrlich, and H. Krieg, A consistent and accurate ab Initio parametrization of density functional dispersion correction (DFT-D) for the 94 Elements H-Pu, J. Chem. Phys. 132, 154104 (2010).

\noindent[13]	J. P. Perdew and A. Zunger, Self-interaction correction to density-functional approximations for many-electron systems, Phys. Rev. B 23, 5048 (1981).